\documentclass[aps,twocolumn,showpacs,superscriptaddress]{revtex4}
\usepackage{epsfig}
\usepackage{graphicx}
\usepackage{amsfonts}
\usepackage[figuresright]{rotating}
\usepackage{amssymb}
\usepackage{amsmath}
\usepackage{subfigure}

\def\avg#1{\langle#1\rangle}

\def\be{\begin{equation}}
\def\ee{\end{equation}}
\def\bea{\begin{eqnarray}}
\def\eea{\end{eqnarray}}

\def\nn{\nonumber}

\begin{document}

\title{ Particle-hole symmetry and interaction effects in the
Kane-Mele-Hubbard model}
\author{Dong Zheng}
\affiliation{State Key
Laboratory of Low-Dimensional Quantum Physics and Department of Physics,
Tsinghua University, Beijing, China 100084}
\affiliation{Department of Physics, University of California, San Diego, 
CA 92093}
\author{Guang-Ming Zhang}
\affiliation{State Key
Laboratory of Low-Dimensional Quantum Physics and Department of Physics,
Tsinghua University, Beijing, China 100084}
\author{Congjun Wu}
\affiliation{Department of Physics, University of California, San Diego,
CA 92093}

\begin{abstract}
We prove that the Kane-Mele-Hubbard model with purely imaginary
next-nearest-neighbor hoppings has a particle-hole symmetry at half-filling.
Such a symmetry has interesting consequences including the absence of charge
and spin currents along open edges, and the absence of the sign problem in
the determinant quantum Monte-Carlo simulations. Consequentially, the
interplay between band topology and strong correlations can be studied at
high numeric precisions. The process that the topological band insulator
evolves into the antiferromagnetic Mott insulator as increasing interaction
strength is studied by calculating both the bulk and edge electronic
properties.
In agreement with previous theory analyses, the numeric simulations show 
that the Kane-Mele-Hubbard model exhibits three phases as increasing
correlation effects: the topological band insulating phase with stable 
helical edges, the bulk paramagnetic phase with unstable edges, and the 
bulk antiferromagnetic phase. 
\end{abstract}

\maketitle

\section{Introduction}

\label{sect:intro} The precise quantization of the Hall conductance in the
integer quantum Hall states is protected by the non-trivial topology of band
structures. This topological property is characterized by the
Thouless-Kohmoto-Nightingale-den Nijs (TKNN) number, or the Chern number
\cite{thouless1982, kohmoto1985}, which takes non-zero values only when
time-reversal symmetry is broken. In recent years, tremendous progress has
been achieved in a new class of topologically non-trivial band insulators in
the presence of time-reversal symmetry, which are termed as topological
insulators \cite{qireview2010,hasan2010,bernevig2006a,qi2008a,
kane2005,sheng2006,moore2007,roy2009,fu2007,fu2007a,zhang2009}. Topological
insulators exist in both two (2D) and three dimensions (3D), which are
characterized by the $Z_{2}$ topological index. These topological states
have robust gapless helical edge modes with odd number of channels in 2D
\cite{kane2005,wu2006,xu2006}, and odd number of surface Dirac cones in 3D
\cite{fu2007,fu2007a,zhang2009}. Topological insulators have been
experimentally observed in 2D quantum wells through transport measurements
\cite{konig2007}, and also in 3D systems of Bi$_{x}$Sb$_{1-x}$, Bi$_{2}$Te$%
_{3}$, Bi$_{2}$Se$_{3}$, and Sb$_{2}$Te$_{3}$ through the angle-resolved
photo-emission spectroscopy \cite{hsieh2008,hsieh2009, xia2009,chen2009},
and the absence of backscattering in the scanning tunneling spectroscopy
\cite{roushan2009,alpichshev2010,zhang_chen_xue2009}.

Interaction effects in topological insulators remain an open question. Due
to their gapped nature, topological insulators remain stable against weak
interactions. However, strong interactions may change their topological
properties. For 2D topological insulators, it has been found that the
two-particle correlated backscattering, which is an interaction effect and
is allowed in the time-reversal invariant Hamiltonian, can gap out the
helical edge states by spontaneous developing magnetic ordering under strong
repulsive interactions \cite{wu2006,xu2006}. In this case, time-reversal
symmetry is spontaneously broken along edges, although the bulk remains
paramagnetic. At mean-field level, interaction effects can destabilize the
quantum anomalous Hall state of the Haldane-Hubbard model \cite{cai2008} and
the 2D topological insulating state of the Kane-Mele-Hubbard (KMH) model
\cite{stephan2010} by developing long-range charge density wave and
antiferromagnetic orders, respectively \cite{stephan2010}. Interactions can
also change the topologically trivial band structures into non-trivial ones
at mean-field level by developing bulk order parameters \cite%
{raghu2008,sun2009,zhang_ran2009, wen2010}. Due to the difficulty of
analytic studies on strong correlation physics, exact results from numeric
simulations are desirable. Recently, an exact diagonalization has been
carried on the spinless Haldane-Hubbard model \cite{varney2010}. A first
order phase transition between quantum anomalous Hall insulating state and
topologically trivial Mott-insulating state is found.

Quantum Monte-Carlo (QMC) simulations play an important role in studying
strongly correlated systems \cite{blankenbecler1981,hirsch1985,
chandrasekharan1999,koonin1997}. A major obstacle to apply the QMC to
fermion systems is the notorious sign problem. In the particular method of
the determinant QMC, the 4-fermion interaction terms are decoupled through
the Hubbard-Stratonovich (HS) transformation and fermions are able to be
integrated out. The resultant fermion determinant, generally speaking, is
not positive-definite, which is the origin of the notorious sign problem.
This problem prevents QMC simulations to achieve a good numerical precision
at low temperatures and large sample sizes. Nevertheless, in a number of
interacting models, the sign problem disappears. As presented in Ref. \cite%
{wu2005}, these models include the negative-$U$ Hubbard model, the positive-$%
U$ Hubbard model at half-filling and in bipartite lattices, and a class of
models whose interactions can be decomposed in a time-reversal invariant way.

We find that the Kane-Mele model augmented by the Hubbard interaction with
purely imaginary next-nearest-neighbor hoppings has a particle-hole
symmetry. Such a symmetry has interesting consequences such as the absence
of edge charge and spin currents, which shows the edge currents are not a
reliable criterion for topological properties. More importantly, the
particle-hole symmetry ensures the absence of the sign problem in the
quantum Monte-Carlo simulations. This provides a wonderful opportunity to
study interaction effects in topological insulating systems. 

In this
article, we perform a determinant QMC study on the stability of the
topological insulating state of the KMH model with the strong Hubbard
interaction $U$. Antiferromagnetic long-range-order has been found at
large values of $U$. Consequently, the quantum phase diagram of the KMH
model can be classified into paramagnetic bulk insulating phases and
antiferromagnetic Mott insulating phases. When further consider the
stability of helical edges with infinitesimal two-particle backscattering,
which is not contained in KMH model but generally allowed by time-reversal
symmetry, the paramagnetic bulk insulating phase can be divided into two
regimes according to their edge state Luttinger parameters \cite{wu2006}. The
topological band insulator with stable helical edges are stable in the weak
interaction regime, while the helical edges become unstable by two-particle
correlated backscattering at the intermediate interaction regime. We have
also studied the nature of spin-liquid phase in the pure Hubbard model with $%
\lambda=0$, showing that it is neither a spontaneous Haldane type quantum
anomalous Hall insulator, nor, a Kane-Mele type quantum spin Hall insulator.

This article is organized as follows. In Section \ref{sect:sign}, we prove
the absence of the sign problem in the KMH model under certain conditions.
In Section \ref{sect:qmc}, we present the simulations on the developing of
antiferromagnetic long-range orders in the bulk. In Section \ref{sect:edge},
the edge properties are studied including both the edge single particle
excitations and the edge spin correlations. In Section \ref{sect:spinliquid}%
, we present the simulation of the charge and spin current orders in the
pure Hubbard model in the honeycomb lattice. Conclusions are given in
Section \ref{sect:conclusion}.


\section{General properties of the KMH model}

\label{sect:sign}

The Kane-Mele model is a straightforward generalization of the Haldane model
in the honeycomb lattice \cite{kane2005} defined as
\begin{eqnarray}
H_0&=&-t\sum_{\langle i,j\rangle,\sigma}c_{i\sigma}^{\dag}c_{j\sigma} +
i\lambda\sum_{\langle\langle i,i^\prime \rangle\rangle\alpha,\beta} \Big\{ %
c^\dagger_{i\alpha} \sigma_{z,\alpha\beta} c_{i^\prime \beta}  \notag \\
&-&c^\dagger_{i^\prime \alpha} \sigma_{z,\alpha\beta} c_{i\beta}\Big\} -\mu
\sum_{i,\sigma} c^\dagger_{i\sigma} c_{i\sigma},  \label{eq:KM}
\end{eqnarray}
where $t$ is the nearest-neighbor (NN) hopping integral as scaled to $1$
below; $\lambda$ is the next-nearest-neighbor (NNN) spin-orbit hopping
integral; $\mu$ is the chemical potential. In the general case of the
Kane-Mele model, the NNN hopping for the spin-$\uparrow (\downarrow)$
electrons are complex-valued and complex-conjugate to each other. As a
special case, the NNN hopping in Eq. \ref{eq:KM} is purely imaginary. The
Hubbard interaction is defined as usual
\begin{eqnarray}
H_{int}=U\sum_{i}\big[n_{i\uparrow}-\frac{1}{2}\big] \big[n_{i\downarrow}-%
\frac{1}{2}\big].  \label{eq:hubbard}
\end{eqnarray}

In this section, we will present the symmetry properties of Eq. \ref{eq:KM}
and Eq. \ref{eq:hubbard}, and prove the absence of the sign problem in the
determinant QMC.


\subsection{Particle-hole symmetry}

Eq. \ref{eq:KM} and Eq. \ref{eq:hubbard} has the particle-hole symmetry at $%
\mu=0$ as explained below. We define the transformation as usual
\begin{eqnarray}
c_{i\sigma }^{\dag }\longrightarrow d_{i\sigma }=(-1)^{i}c_{i\sigma
}^{\dag}, \ \ \ c_{i\sigma }\longrightarrow d_{i\sigma }^{\dag
}=(-1)^{i}c_{i\sigma }.  \label{eq:p-h}
\end{eqnarray}
Under this transformation, a Hermitian fermion bilinear operator connecting
two sites belonging to two different sublattices transforms as
\begin{eqnarray}
c^\dagger_{i\sigma} K_{ij} c_{j\sigma}+ c^\dagger_{j\sigma} (K_{ij})^*
c_{i\sigma} \longrightarrow  \notag \\
d^\dagger_{i\sigma} (K_{ij})^* d_{j\sigma}+ d^\dagger_{j\sigma} K_{ij}
d_{i\sigma},  \label{eq:p-h-NN}
\end{eqnarray}
while that connecting two different sites in the same sublattice transforms
as
\begin{eqnarray}
&&c^\dagger_{i\sigma} K_{ii^\prime} c_{i^\prime\sigma} +
c^\dagger_{i^\prime\sigma} (K_{ii^\prime})^* c_{i^\prime\sigma}
\longrightarrow  \notag \\
&-&d^\dagger_{i\sigma} (K_{ii^\prime})^* d_{i^\prime\sigma} -
d^\dagger_{i^\prime\sigma} K_{ii^\prime} d_{i^\prime\sigma}.
\label{eq:p-h-NNN}
\end{eqnarray}
The onsite particle density transforms as
\begin{eqnarray}
c^\dagger_{i\sigma}c_{i\sigma}-\frac{1}{2} \longrightarrow \frac{1}{2}%
-d^\dagger_{i\sigma}d_{i\sigma},  \label{eq:density}
\end{eqnarray}
where no summation over spin-index is assumed in Eq. \ref{eq:density}.
Clearly in Eq. \ref{eq:KM}, the NN-hopping is real and the NNN-hopping is
purely imaginary, thus its band structure is invariant at $\mu=0$. Eq. \ref%
{eq:hubbard} is obviously invariant. The particle-hole symmetry also implies
that $\mu=0$ corresponds to half-filling.


\subsection{Absence of the charge and spin currents}

An important conclusion based on the particle-hole symmetry is that both
charge and spin currents vanish on all the bonds for the KMH model of Eq. %
\ref{eq:KM} and Eq. \ref{eq:hubbard} at $\mu=0$. This result applies to
arbitrary boundary conditions with broken bonds but with the homogeneous
on-site potential which maintains the particle-hole symmetry on each site.
The proof is straightforward. Through the continuity equation, the current
operators of each spin component along the NN and NNN bonds are defined as
\begin{eqnarray}
J^{NN}_{i j,\sigma}= i t \big(c^\dagger_{i\sigma} c_{j\sigma} -
c^\dagger_{j\sigma} c_{i\sigma}\big),  \notag \\
J^{NNN}_{i i^\prime,\sigma}= \lambda \big(c^\dagger_{i\sigma}
c_{i^\prime\sigma} +c^\dagger_{i^\prime\sigma} c_{i\sigma}\big),
\label{eq:nnn_curr}
\end{eqnarray}
respectively, where no summation over spin-index is assumed. Both $J^{NN}$
and $J^{NNN}$ are odd under the particle-hole transformation, thus they
vanish even with the open-boundary condition. By the same reasoning, the
charge current also vanishes in the Haldane-Hubbard model with the purely
imaginary NNN-hoppings and the particle-hole symmetric charge interactions
of
\begin{eqnarray}
H_{NN,int}=\sum_{ij} V_{ij} (n_i-\frac{1}{2}) (n_j-\frac{1}{2}).
\end{eqnarray}

This result shows that edge charge and spin currents are not good criteria
for quantum anomalous Hall and topological insulators. In order to have a
better understanding on this counter-intuitive result, we have considered
the simplest non-interacting Haldane model with the purely imaginary NNN
hoppings by diagonalization. There are indeed gapless one-dimensional single
particle chiral edge modes clearly seen from the spectra as commonly
presented in literatures. Clearly this branch of edge mode contributes to
edge currents. However, we find that the continuous bulk spectra also
contribute to edge currents. Perfect cancellation occurs which results in
zero current on each bond, including each edge bond, although we know for
sure that the band structure is topologically non-trivial. For interacting
models, there are no well-defined single particle states. We cannot separate
the edge and bulk contributions anymore. Nevertheless, we expect that
current correlation functions should exhibit difference between topological
insulators and trivial insulators.

Another conclusion inferred from the particle-hole symmetry is that the
average particle density for each spin component on each site is strictly $%
\frac{1}{2}$ even when the translational symmetry is broken. For example, it
applies to any disordered pattern of the hopping integrals, as long as the
NN hoppings are real and the NNN hoppings are purely imaginary.

Edge currents do appear if the particle-hole symmetry is broken. For
example, for the non-interacting Haldane model with generally complex-valued
NNN hoppings, edge currents appear along open boundaries. So far we only
consider the sharp edges of broken bonds but with homogeneous on-site
potential. For edges with the confining single particle potential, the
particle-hole symmetry is broken which also results in edge currents. In
particular, for a weak linear external potential, the linear response should
still give rise to quantized Hall conductance in the insulating region.


\subsection{Absence of the QMC sign problem}

The Hubbard model on the honeycomb lattice, which corresponding to the case
of $\lambda=0$ of Eq. \ref{eq:KM} and Eq. \ref{eq:hubbard} has been recently
simulated at half-filling \cite{meng2010}. As Hubbard $U$ increases from
zero to a moderate value and then the strong coupling regime, the ground
state emerges from a semi-metal phase, to a new spin-liquid phase and then
the antiferromagnetic insulating phase. Below we will prove that the sign
problem still vanishes with nonzero values of $\lambda$.
The absence of the sign problem can be proved for both the finite temperature 
and the zero temperature algorithms for the determinant QMC.
In this subsection, we prove this property for the finite temperature
method for simplicity, and leave the more lengthy proof for the zero 
temperature algorithm in Appendix \ref{sect:appendix}.
We emphasize that the simulations presented in this article are
done at the zero temperature.

Just as the Ref.(\cite{meng2010}) does, we employ a discrete HS
transformation which respects the $SU(2)$ symmetry for every fixed HS field
configuration by decoupling in the density channel. We rewrite the Hubbard
interaction and decompose it in the density channel by using imaginary
numbers as
\begin{eqnarray}
e^{-\Delta U(n_{\uparrow }+n_{\downarrow }-1)^{2}/2}&=&\sum_{l=\pm 1,\pm
2}\gamma _{i}(l)e^{i \eta _{i}(l)\sqrt{\Delta \tau \frac{U}{2}}(n_{\uparrow
}+n_{\downarrow }-1)}  \notag \\
&+&\mathcal{O}(\Delta \tau ^{4}).
\label{eq:HS}
\end{eqnarray}
where the discretized HS fields take values of $\gamma (\pm 1) =1+\sqrt{6}/3$%
, $\gamma (\pm 2) =1-\sqrt{6}/3$; $\eta (\pm 1)=\pm \sqrt{2(3-\sqrt{6})}$,
and $\eta (\pm 2)=\pm \sqrt{2(3+\sqrt{6})}$.

For the convenience of presentation, we prove the absence of the sign
problem in the finite temperature formalism with $\beta=1/T$. The proof for
the zero temperature projector algorithm is similar. The partition function
at half-filling reads
\begin{widetext}
\bea
Z &=& \sum_{\{ l \}} \Big\{ \left( Tr\prod_{p=M}^{1}
e^{-\Delta \tau \sum_{i,j}c_{i\uparrow }^{\dag } K_{ij}^{\uparrow } c_{j\uparrow }}
e^{i\sqrt{\Delta \tau U/2} \sum_{i} \eta_{i,p}(l)
(c_{i\uparrow }^{\dag }c_{i\uparrow}-\frac{1}{2})} \right) \nn \\
&\times&
\left( Tr\prod_{p=M}^{1}e^{-\Delta \tau \sum_{i,j}c_{i\downarrow}^{\dag }
K_{ij}^{\downarrow }c_{j\downarrow }}e^{i\sqrt{\Delta \tau U/2}
\sum_{i} \eta_{i,p}(l) (c_{i\downarrow }^{\dag }c_{i\downarrow }
-\frac{1}{2})}\right) \prod_{i,p}\gamma _{i,p}(l)\Big\},
\eea
\end{widetext}
where $\sum_{\{l\}}$ sums over all the configurations of the discrete HS
fields $\eta_{i,p}(l)$ and $\gamma_{i,p}(l)$; $i$ and $p$ are indices of
discretized grids along the spatial and temporal directions, respectively; $%
Tr$ takes the trace of the fermion space; $\Delta \tau$ is the discretized
time slice which is set to $0.05$ in the simulations in this article; $M
\Delta \tau$ equals the imaginary time $\beta$. By using the particle-hole
transformation defined in Eq. \ref{eq:p-h}, we show that the onsite particle
density transforms according to Eq. \ref{eq:density}; the NN-hopping matrix
kernel transforms according to Eq. \ref{eq:p-h-NN}; the NNN-hopping matrix
kernel transforms according to Eq. \ref{eq:p-h-NNN}.

When the following two conditions are satisfied, the fermion determinants of
two spin components are complex-conjugate to each other, thus the product of
them is positive-definite:
\begin{eqnarray}
K_{ij}^{\sigma }&=&\left( K_{ji}^{\bar{\sigma}}\right) ^{\ast }=
K_{ij}^{\bar\sigma} \ \ \ \mbox {for NN-hopping};  \notag \\
K_{ij}^{\sigma }&=&-\left( K_{ji}^{\bar{\sigma} }\right) ^{\ast
}=-K_{ij}^{\bar\sigma} \ \ \ \mbox {for NNN-hopping.}
\label{eq:nosign}
\end{eqnarray}
Apparently, Eq. \ref{eq:KM} and Eq. \ref{eq:hubbard} satisfy these
conditions, and thus are sign problem free.

Please note that the KMH mode is sign problem free only when the NNN-hopping
is purely imaginary. Generally speaking, the interacting model without the
sign problem can have complex-valued hoppings with opposite signs, which
still gives rise to opposite Chern numbers for the band structures of spin-$%
\uparrow$ and $\downarrow$, respectively. However, they are not related by
time-reversal symmetry anymore.

\section{The QMC study on the bulk properties of the KMH-model}
\label{sect:qmc}

The Hubbard model in the honeycomb lattice, which corresponds the case of $%
\lambda=0$ in Eq. \ref{eq:KM} and Eq. \ref{eq:hubbard}, has been simulated
in Ref. \cite{meng2010}. When $U$ increases from zero, the single particle
charge gap appears at $U=3.7$, while the antiferromagnetic long-rang order
emerges at $U=4.3$. The mismatch reveals an exotic spin liquid phase in
between. When the intrinsic spin-orbit coupling, \textit{i.e.}, the NNN
hopping term in Eq. \ref{eq:KM}, enters, the model describes the topological
band insulator. It already has a band gap even at $U=0$. As increasing $U$,
the antiferromagnetic structure factor is still a good quantity to tell when
the magnetic long range order appears. However, the bulk gap is no longer an
appropriate quantity to judge a possible transition from the topological
band insulator to a antiferromagnetic Mott-insulator. Here we use the local
single particle gap on edge sites as an indicator of the stability of edge
states and topological properties. We also study the edge effects to
antiferromagnetic correlations. In this section, we will simulate the bulk
antiferromagnetic structure factor, and leave the study of edge properties
in Section \ref{sect:edge}.


\subsection{Sampling parameters of our simulations}

Based on the above proof of the absence of the sign problem, we perform the
QMC simulation for the KMH model at {\it zero temperature}
by using the projective
method \cite{assaad2008}. We perform measurements from $10$ different random
number series and each independent measurement has $500$ sample sweeps after
warming up, the discrete imaginary time step $\Delta\tau$ is set to be $0.05$%
. In this section, we use periodic boundary conditions for bulk properties
calculation, e.g., the bulk antiferromagnetic structure factor.


\subsection{The developing of the bulk antiferromagnetic long range order}

\label{sect:antiferro} The spin-orbit NNN hopping in Eq. 1 breaks the $SU(2)$
symmetry but preserves the conservation of $S_z$. As a result, the
antiferromagnetic correlation of $S_z$ should be different from those of $%
S_x $ and $S_y$. In the large $U$-limit, the NNN hopping generates an
anisotropic exchange as
\begin{eqnarray}
H_{ex,NNN}=-J^\prime (S^x_i S^x_{i^\prime} +S^y_i S^y_{i^\prime}-S^z_i
S^z_{i^\prime})
\end{eqnarray}
with $J^\prime=4\lambda^2/U$, which is ferromagnetic in the $xy$-plane and
antiferromagnetic along the $z$-direction \cite{stephan2010}. As the
combined effect from the NNN anisotropic exchange and NN isotropic
antiferromagnetic exchange, the magnetic exchange along the $z$-axis is
frustrated while those along $x$ and $y$- axes are not. Thus the Neel
ordering favors the easy $xy$-plane.

\begin{figure}[tbp]
\centering
{\epsfig{file=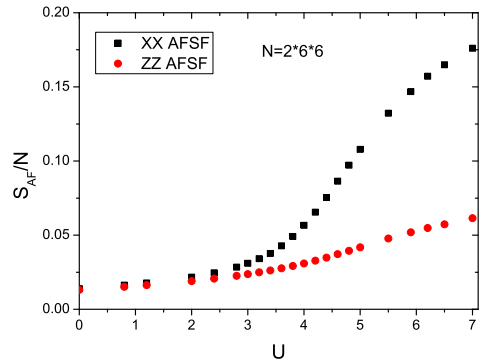,clip=1,width=0.9\linewidth,angle=0}}
\caption{(Color online) The comparison between the antiferromagnetic
structure factors $S^{zz}_{AF}$ along the $z$-axis and $S^{xx}_{AF}$ in the $%
xy$-plane at $\protect\lambda=0.1$ for the size of $N=2\times L \times L$
with $L=6$. The easy-plane feature is clear. }
\label{fig:xxzzafsf}
\end{figure}

\begin{figure}[tbp]
\centering
{\epsfig{file=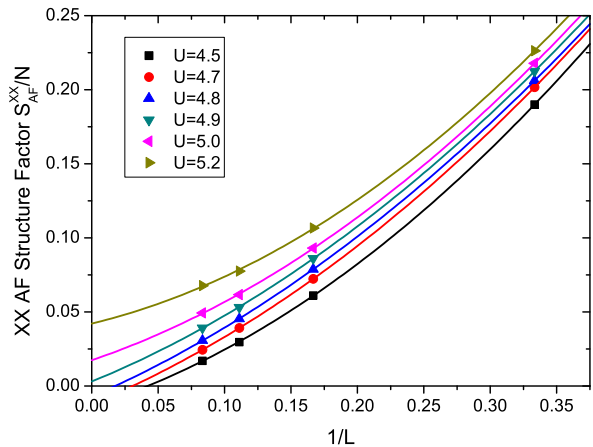,clip=1,width=0.9\linewidth,angle=0}}
\caption{(Color online) The finite-size scaling of the $xx$%
-antiferromagnetic structure factors calculated at $\protect\lambda=0.1$ for
the sizes of $N=2\times L\times L$ ($L=3,6,9$ and $12$), and the different
values of $U$ indicated in the inset. Finite values of $S^{xx}_{AF}/N$ in
the thermodynamic limit appear at $U\geq U_c$ with $U_c\approx 4.9$. }
\label{fig:xxafsf}
\end{figure}

Our QMC simulations have confirmed this picture. The antiferromagnetic
structure factor along the $x$-direction ($xx$-AFSF) and the $z$-direction ($%
zz$-AFSF) are defined as
\begin{eqnarray}
S_{AF}^{xx}&=&\frac{1}{N}\langle G|\left[ \sum_{i}(-1)^{i}S_{i}^{x} \right]
^{2}|G\rangle,  \notag \\
S_{AF}^{zz}&=&\frac{1}{N}\langle G|\left[ \sum_{i}(-1)^{i}S_{i}^{z} \right]
^{2}| G\rangle,
\end{eqnarray}
where $\langle G|..|G\rangle$ means average over the ground state; $%
N=2\times L \times L$ is the number of sites; $L$ is the size; $(-)^i$ takes
the values of $\pm 1$ for the $A$ and $B$-sublattices, respectively. The
comparison between $S_{AF}^{xx}$ and $S_{AF}^{yy}$ is plotted in Fig. \ref%
{fig:xxzzafsf}, which clearly shows the easy-plane feature.

\begin{figure}[tbp]
\centering
{\epsfig{file=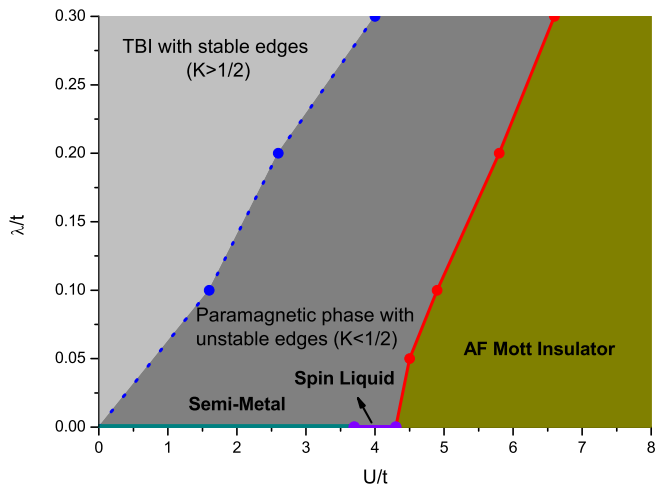,clip=1,width=\linewidth,angle=0}}
\caption{(Color online) The QMC simulation of the phase diagram of the KMH
model. The antiferromagnetically long-range ordered phase appears at strong
correlation regime. The paramagnetic phase is divided into two regimes:
topological band insulator (TBI) with stable helical edges, and bulk
paramagnetic phase with unstable edges (see further discussions in Sect.
\protect\ref{sect:stability}). The two critical values of $U$ at $\protect%
\lambda=0$ are from Ref. \protect\cite{meng2010} by Meng \textit{et al.},
which are also confirmed in our QMC simulations.}
\label{fig:phase_diagram}
\end{figure}

Below we will use the $xx$-AFSF to describe the antiferromagnetic
properties, and perform the simulation at $\lambda=0.1$ with different
values of $U$ and sample sizes of $L=3,6,9,12$. The extrapolation to the
thermodynamic limit for different Hubbard $U$ is plotted in of Fig. \ref%
{fig:xxafsf}. It can be seen that the magnetic long range order emerges at $%
U_c=4.9\pm 0.1$ for $\lambda=0.1$. In Fig. \ref{fig:phase_diagram} we
present the QMC simulation on the magnetic phase diagram of the KMH model
for in the parameter space of $(U, \lambda)$. The phase boundary separating
the AF long-range-ordered phase and non-magnetic phases are marked for
various values of $\lambda$. The spin-orbit coupling opens the band gap at
the order of $\lambda$, thus the interaction effect $U$ becomes important
only when $U$ is larger than $\lambda$. As a result, the critical value of $%
U_c$ for the onset of the AF phase increases with $\lambda$.

The phase diagram Fig. \ref{fig:phase_diagram} exhibits a large regime of
non-magnetic insulating state outside the AF phase at $\lambda\neq 0$. At
small values of $U$, it should be the $Z_2$ topological band insulating
phase which is stable against weak interactions. 
As increasing $U$, it enters the AF Mott
insulating phase at a critical line of $U_c$. In an updated version of Ref.
\cite{hohenadler2010}, it is found that the spin-liquid phase also extends
to a small but finite value of $\lambda$. However, the nature of this
spin-liquid state remains unclear. The bulk paramagnetic regime actually has
rich internal structures. According to the stability of the helical edge
states with respect to the two-particle spin-flip backscattering, this
paramagnetic insulating phase is divided into two different regimes with the
effective edge Luttinger parameter $K<(>)\frac{1}{2}$, respectively. The
analysis is presented below in Sect. \ref{sect:stability}.


\section{The QMC study of the edge properties of the KMH model}

\label{sect:edge}

\begin{figure}[tbp]
\centering
{\epsfig{file=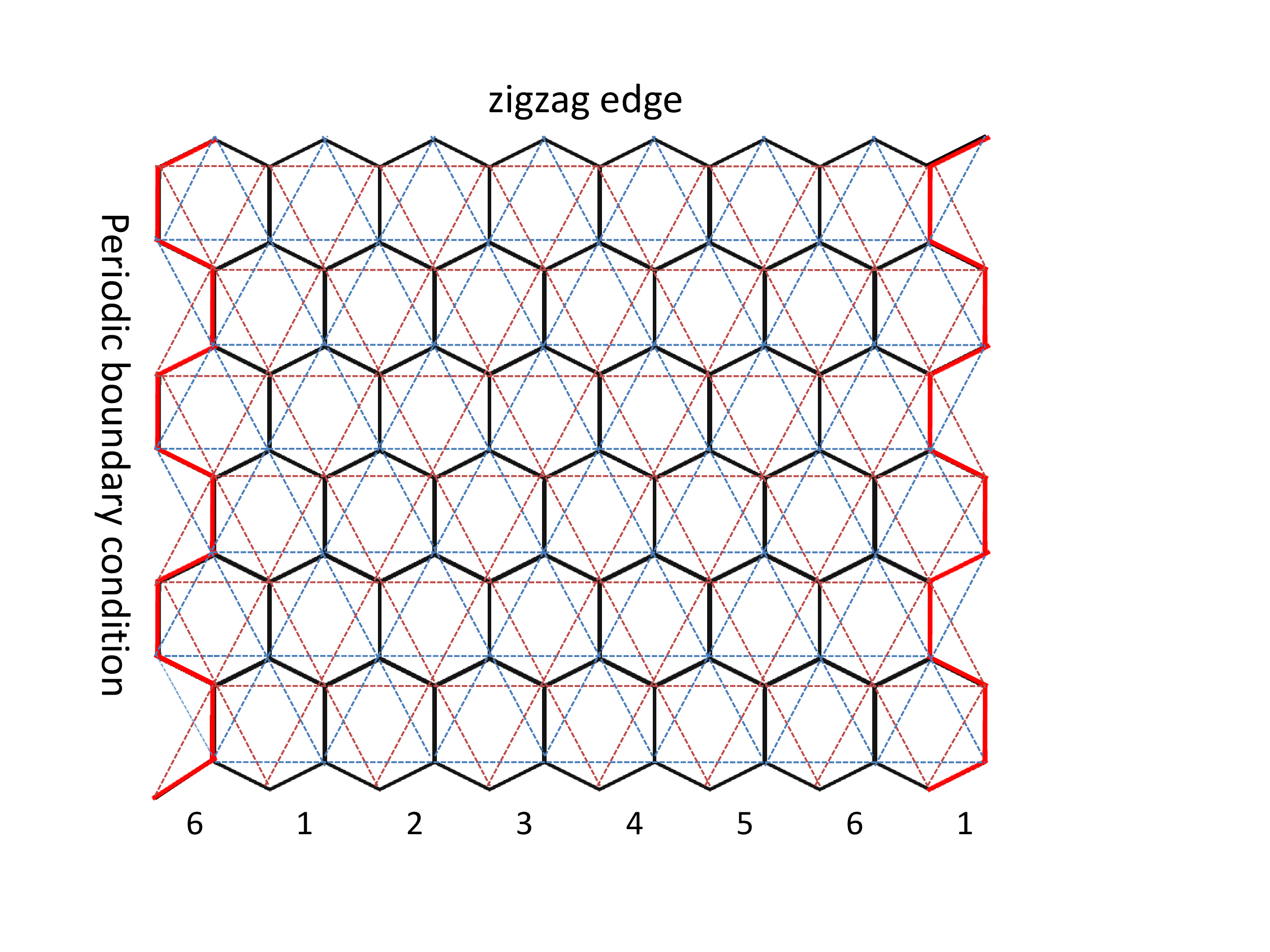,clip=1,width=0.8\linewidth,angle=0}}
\caption{(Color online) The KMH model in the lattice with the zig-zag edges.
The boundary conditions are periodical and open along the $x$ and $y$%
-directions, respectively. The NNN-bonds between two closest tips on the
zig-zag edges are removed. }
\label{fig:open}
\end{figure}

We believe that the edge properties is crucial to expose the topological
aspect of the KMH model. In this section, we will show that the
antiferromagnetic correlations along the edge become strongly relevant as
increasing the Hubbard $U$ while the bulk remains paramagnetic. We consider
the lattice configuration plotted in Fig. \ref{fig:open} with the periodical
and open boundary conditions along the $x$ and $y$-directions, respectively.

\begin{figure}[tbp]
\centering
{\epsfig{file=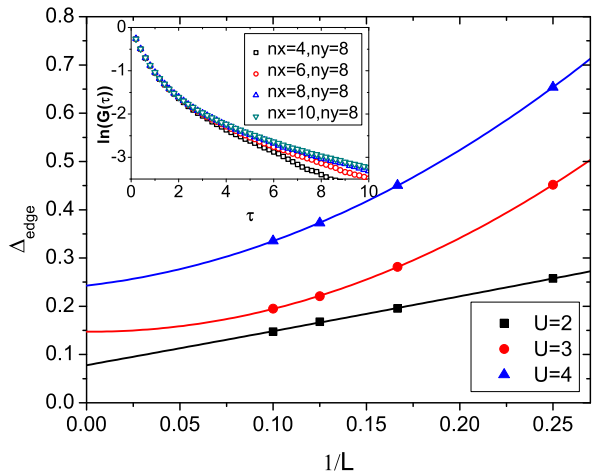,clip=1,width=0.9\linewidth,angle=0}}
\caption{(Color online) The extrapolation of local single particle gap for
the tip sites on the zig-zag edges of a ribbon geometry with the size of $%
2\times L \times n_y$ with $n_y= 8$. In the inset, the logarithms of onsite
time-displaced Green's functions $\ln G(i,i;\protect\tau)$ of the tip sites
is depicted for $U=2$. The slopes of the long time tails measure the edge
excitation gap $\Delta_{edge}$. Here $\protect\lambda$ is set to be $0.1$ in
this calculation.
Due to the one-dimensional nature of the edge, its densities of states 
in the thermodynamic limit are depleted according to the power-law
determined by the Luttinger parameter, and thus are
difficult to be distinguished from finite gaps.
}
\label{fig:edge_gap}
\end{figure}


\subsection{The single-particle excitations}

As proved in Sect. \ref{sect:sign}, the edge currents, both for charge and
spin, are always zero due to the particle-hole symmetry. We use another
quantity, the local single particle excitation gap on edge sites, to check
whether the edges are gapped or gapless. It can be extracted from the tail
of on-site time displaced Green's function on the edge $\ln G(i,i;\tau)
\sim\Delta_{edge}\tau$, which is defined by
\begin{eqnarray}
G(i,i;\tau)=\frac{1}{L}\langle G|\sum_{i\in tip}c_{i\uparrow}^{\dag}
(\tau)c_{i\uparrow}(0)+c_{i\downarrow}^{\dag}(\tau)c_{i\downarrow}(0)|G%
\rangle,  \notag \\
\end{eqnarray}
where $|G \rangle$ is the many-body ground state. The dependency of $\ln
G(i,i;\tau)$ with $\tau$ for the site $i$ on the tip of the zig-zag edges
are plotted in the inset of Fig. \ref{fig:edge_gap}, where the long tail of $%
\ln G(i,i;\tau)$ shows a linear behavior with $\tau$ and the slope measures
the excitation gap. Here the lattice has a ribbon geometry with $n_y$
zig-zag rows. We fix the width of the ribbon $n_y=8$ and increase its
length. The extrapolations of the edge excitation gaps with $L$ are depicted
in Fig. \ref{fig:edge_gap} with $\lambda$ fixed at 0.1 and different values
of $U<U_c$. Clearly increasing $U$ significantly reduces the weight of the
low energy spectra.

The bosonization analysis of the stability of the helical edge states has
been performed in Ref. \cite{wu2006,xu2006}. For the parameter regime of
Fig. \ref{fig:edge_gap}, the bulk remains paramagnetic, or, time-reversal
invariant. For the current KMH-model, $S_z$ is conserved which prohibits the
existence of the two-particle spin-flip scattering term to open the gap. The
Luttinger liquid theory of such a helical edge branch, \textit{i.e.}, the
right and left movers are with opposite spin polarizations, is characterized
by only one Luttinger parameter $K$, which describes the forward scattering
between these two branches. Due to the helical nature of the edge states,
the long wavelength charge fluctuations and the $z$-component of the spin
fluctuations are not independent but are conjugate to each other. Both of
them are gapless in the thermodynamic limit, and so does the single particle
edge excitations. The onsite imaginary time single-particle Green's function
decays as $1/\tau^\alpha$ with the exponent
\begin{eqnarray}
\alpha=K+1/K.
\end{eqnarray}
At $K\ll 1$, the low energy density of states does not open a full gap but
are depleted according to a power-law, and thus exhibit a pseudo-gap
behavior. The non-zero gap values in Fig. \ref{fig:edge_gap} may be an
artifact of finite size scaling and a result of tunneling between two
opposite edges. A more detailed numerical analysis is needed to further
clarify the nature of the single particle excitations.


\subsection{Edge spin structure factors}

\begin{figure}[tbp]
\centering
{\epsfig{file=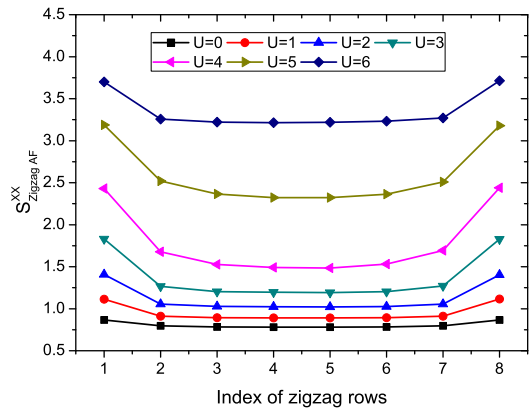,clip=1,width=0.8\linewidth,angle=0}}
\caption{(Color online) The row $xx$-AFSF defined in Eq. \protect\ref%
{eq:lines_af} for each zig-zag rows parallel to the boundary. The parameters
values are $\protect\lambda=0.1$; the sample size $2\times L \times L$ with $%
L=8$; and different values of $U$ indicated in the inset. The row-indices 1
and 8 correspond to the boundary rows, and those of 4 and 5 corresponds to
the central rows. }
\label{fig:lines_af}
\end{figure}

We further investigate the edge effects to the antiferromagnetic
correlations. We define the antiferromagnetic structure form factor for each
zig-zag row parallel to the zig-zag boundary as
\begin{eqnarray}
S_{Zigzag,AF}^{xx}(m)&=&\frac{1}{2L}\langle G |[\sum_{i} (-1)^{i}S_{m,
i}^{x}]^2| G\rangle,  \label{eq:lines_af}
\end{eqnarray}
where $m$ is the index of the zig-zag row; $i$ is the site index along the $%
m $-th zig-zag line; $2L$ is the number of sites in each row. The $xx$-AFSF
for all the rows are depicted in Fig. \ref{fig:lines_af}.

It is interesting to observe that the AF correlations are strongest on
edges, and become weaker inside the bulk. This effect is most prominent at
small and intermediate values of $U$, because the single particle band gap
due to $\lambda$ is suppressed around edges, which enhances the interaction
effects. When $U\geq U_c\approx 4.9$, the bulk antiferromagnetism develops.
The antiferromagnetic correlations along both the edge and central rows are
enhanced by $U$. However, their difference is suppressed due to the
disappearance of the helical edge states.

\begin{figure}[tbp]
\centering
{\epsfig{file=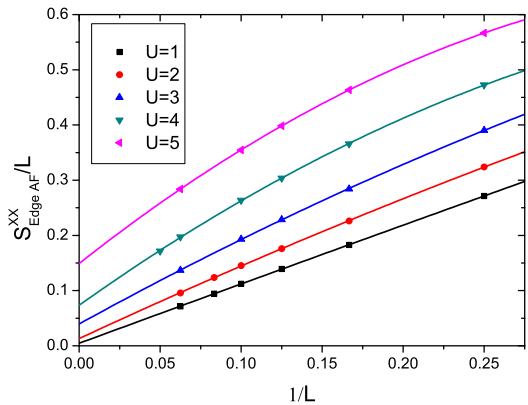,clip=1,width=0.8\linewidth,angle=0}}
\caption{(Color online) The finite size scaling of the $xx$-AFSF defined in
Eq. \protect\ref{eq:lines_af} for the edge row with $\protect\lambda=0.1$.
The size of this ribbon is $2\times L\times 4$. We emphasize that due to the
1D nature of the edge and the $U(1)$ spin symmetry, this scaling actually
shows the power-law correlation rather than the true-long-range order. The
finite intercepts are mainly due to small size effects.}
\label{fig:edge_af}
\end{figure}

The finite-size scaling of the $xx$-AFSF for the edge rows for different
values of $U$ are presented in Fig. \ref{fig:edge_af}. Compared with the $xx$%
-AFSF calculated in the bulk (Fig. \ref{fig:xxafsf}), the edge
antiferromagnetic correlations are much stronger than those of the bulk.
Although the extrapolation to the infinite size in Fig. \ref{fig:edge_af}
implies a finite value of the Neel order of $S_x$ on the edge, we believe
that it is an artifact due to the power-law scaling of the AF correlations.
The 1D nature of the edge states and the conservation of $S_z$ prohibits the
true long range Neel ordering of $S_{x,y}$ but allows the quasi-long-range
ordering, which is confirmed in the two-point spin correlations in Sect. \ref%
{sect:stability}.

\subsection{The stability of the helical edges}

\label{sect:stability}

\begin{figure}[tbp]
\centering
{\epsfig{file=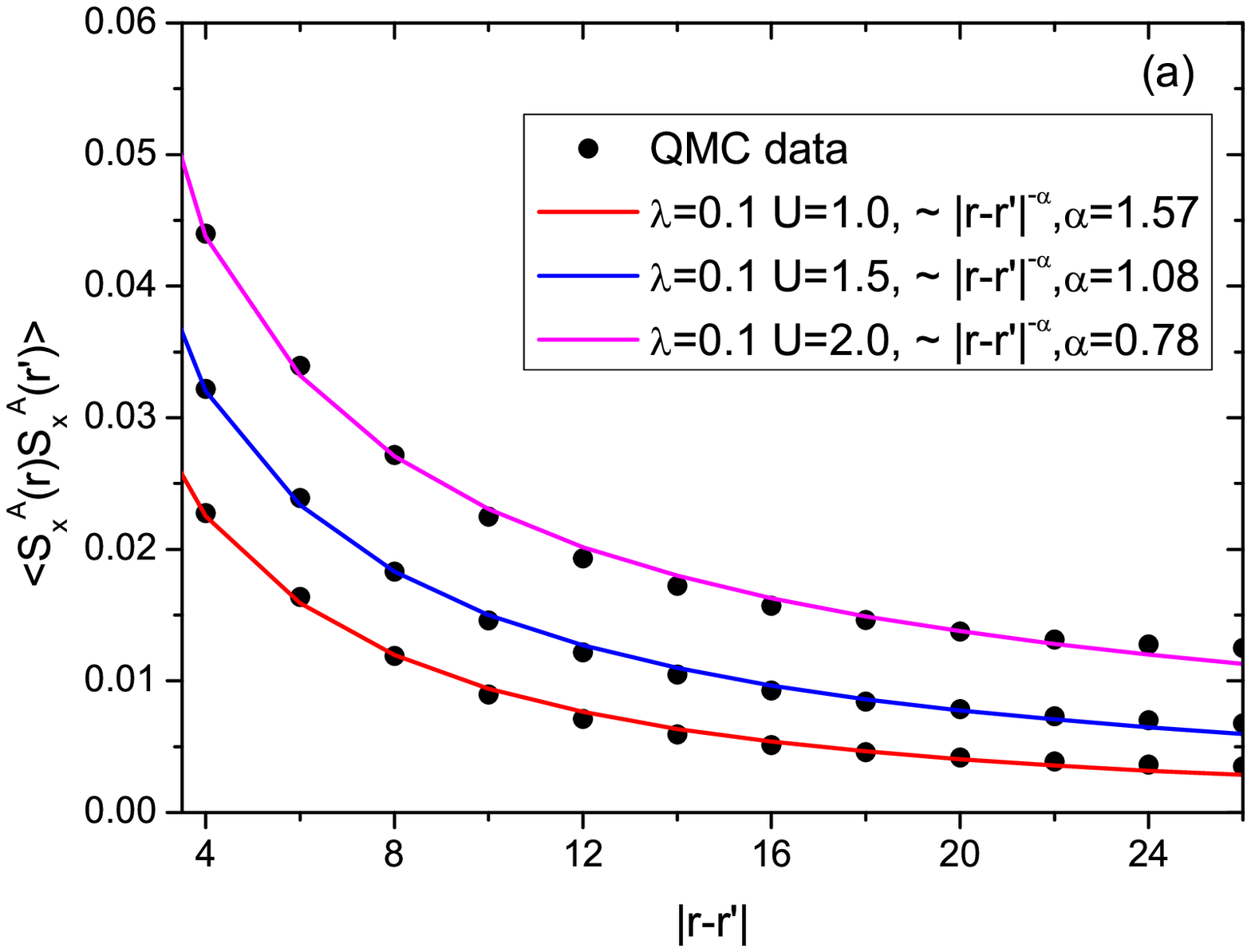,clip=1,width=0.8\linewidth} %
\epsfig{file=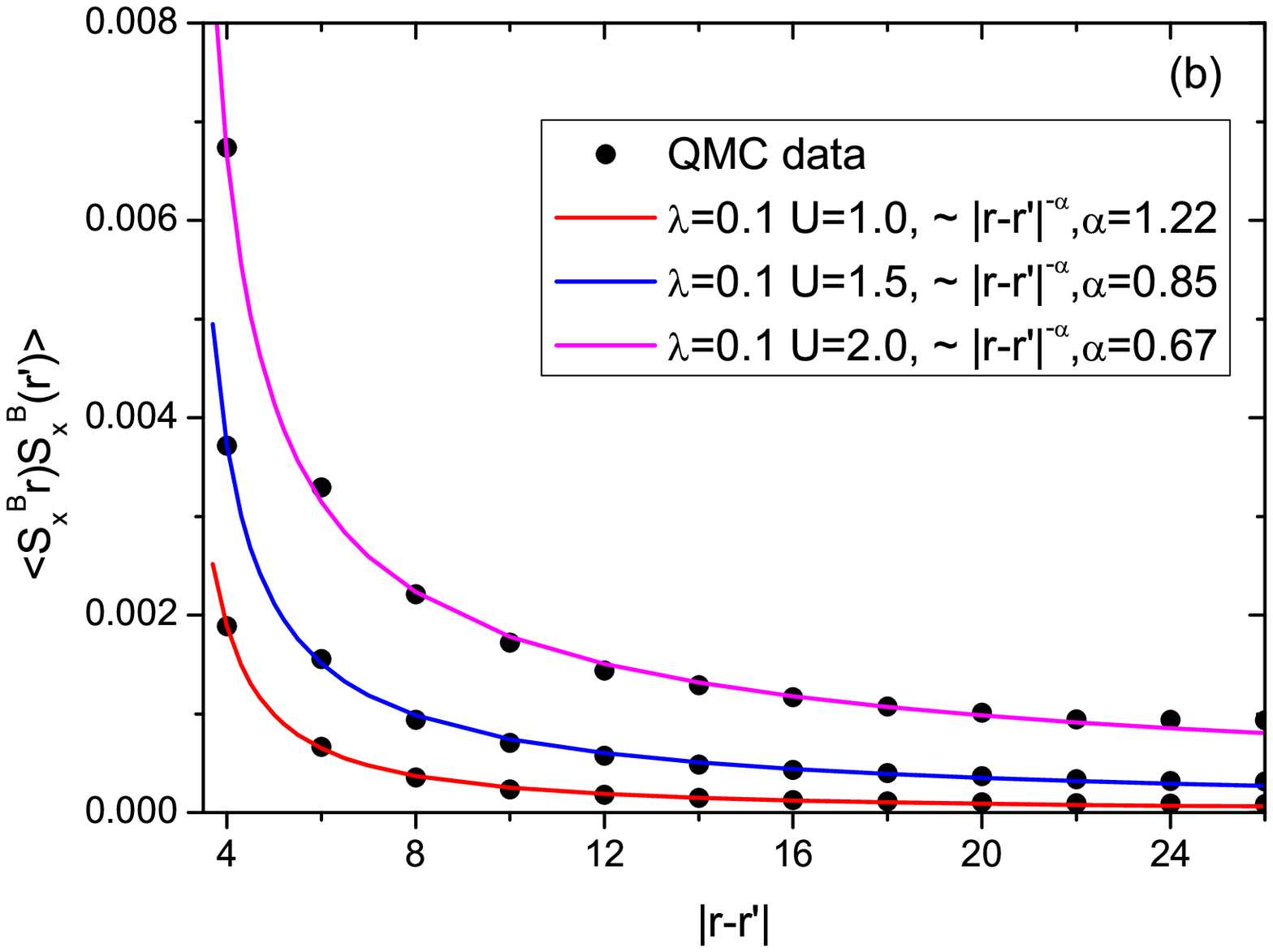,clip=1,width=0.8\linewidth} %
\epsfig{file=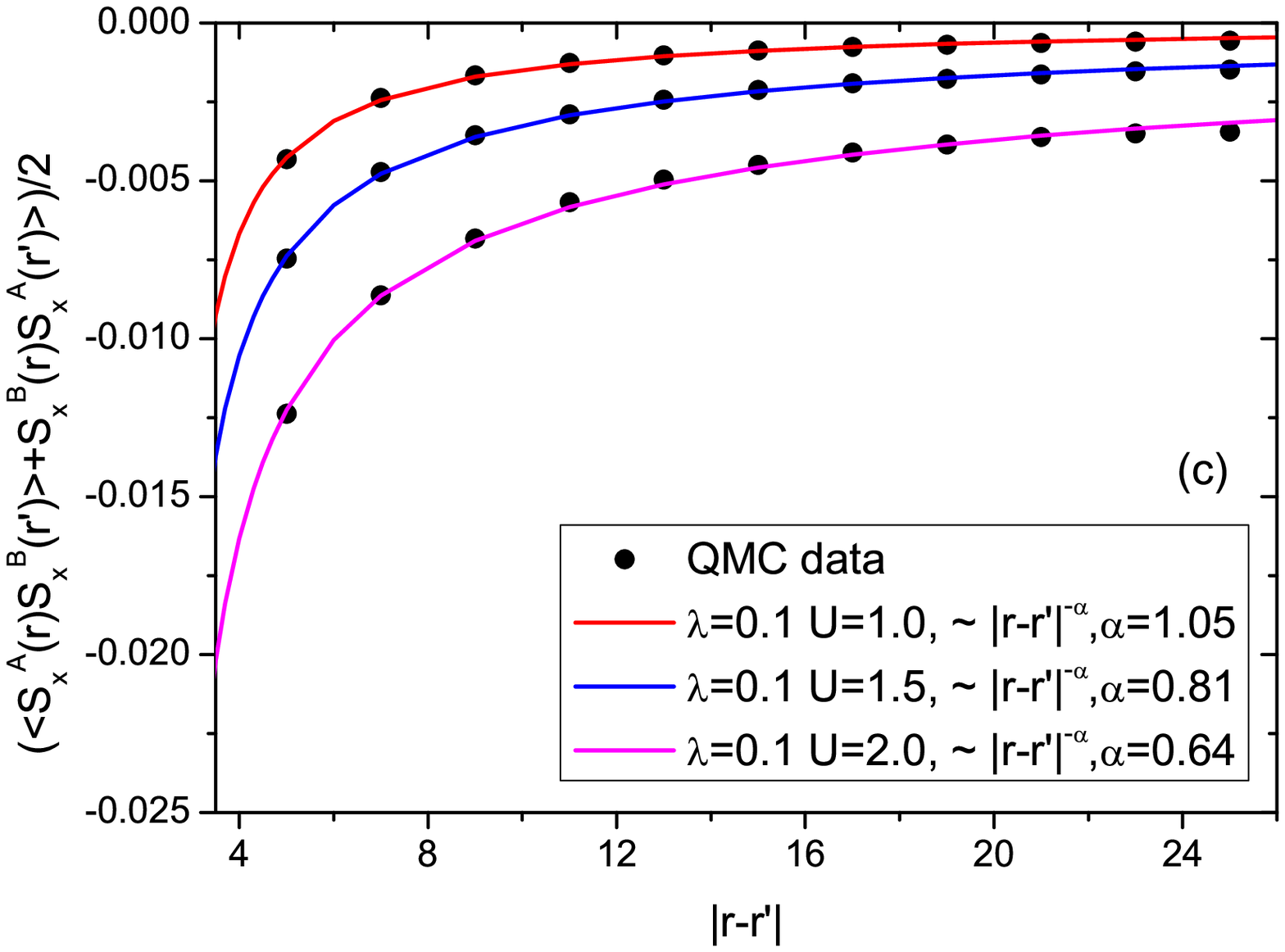,clip=1,width=0.8\linewidth} }
\caption{(Color online) The two-point equal-time spin correlation functions
along the zig-zag edge with $\protect\lambda=0.1$ at values of $U$ denoted
in the insets. The sizes of the ribbon is $2\times 34 \times 4$. Because the
zig-zag edge contains the sites of both $A$ and $B$ type, three different
types of correlations are plotted in (a), (b), and (c), respectively. The
Luttinger parameters are fitted from the correlation among $A$-sites on the
tips as $K\approx 0.8,0.5$ and $0.4$ for $U=1, 1.5$ and $2$, respectively. }
\label{fig:edgeaf_power}
\end{figure}

According to the bosonization analysis in Ref. \cite{wu2006}, the scaling
dimension of the $2k_f$ Neel order of the $xy$-components is $K$, thus their
equal-time correlations decays as $1/|x-x^\prime|^{2K}$. If the condition of
the conservation of $S_z$ is released, a time-reversal invariant
two-particle correlated spin-flip backscattering term is allowed as
\begin{eqnarray}
H_{bg,2pct}=\int dx ~\psi^\dagger_{R\uparrow}\partial_x
\psi^\dagger_{R\uparrow} \psi_{L\downarrow}\partial_x\psi_{L\downarrow}+h.c.
\label{eq:2pctbg}
\end{eqnarray}
At the particle-hole symmetric point of the KMH-model that we are
simulating, the above term becomes the Umklapp term which conserves the
lattice momentum. Such a term reduces the $U(1)$ spin symmetry down to $Z_2$%
. It has the scaling dimension $4K$, and becomes relevant at $K<K_c=1/2$. In
this case, it opens a gap by developing the long range $2k_f$ magnetic
ordering of $S_x$ or $S_y$. Even for the cases that the two-particle
spin-flip backscattering are random disordered or at a single site, they
still can destabilize the helical edge states at smaller values of the
Luttinger parameter $K$ \cite{wu2006}.

According to the above analysis, the bulk paramagnetic regime at weak and
intermediate coupling strengths should be divided into two regimes. At weak
interactions, the helical edge states are stable against interaction
effects. The two-particle backscattering terms only have perturbative
effects. On the other hand, at intermediate level of interaction strength,
interaction effects are non-perturbative which breaks time-reversal symmetry
along edges and thus destroys the helical edges. We emphasize that this
destabilizing helical edges occurs when the bulk remains paramagnetic and
time-reversal invariant.

To numerically verify this picture, we present the calculation of the real
space equal-time two-point correlations along the zig-zag edge in Fig. \ref%
{fig:edgeaf_power}. Since each unit cell contains two non-equivalent sites,
we denote the sites on the tips of the edge as $A$-sites and the other
slightly inner sites as $B$-sites. The correlation functions are defined as
\begin{eqnarray}
C_{AA}(r,r^\prime)&=&\langle G| S^A_x(\vec r) S^A_x(\vec r^\prime)|G\rangle,
\notag \\
C_{BB}(r,r^\prime)&=&\langle G| S^B_x(\vec r) S^B_x(\vec r^\prime)|G\rangle,
\notag \\
C_{AB}(r,r^\prime)&=&\frac{1}{2} \big\{\langle G| S^A_x(\vec r) S^B_x(\vec
r^\prime)|G\rangle  \notag \\
&+&\langle G| S^B_x(\vec r) S^A_x(\vec r^\prime)|G\rangle \big\},
\end{eqnarray}
where $\vec r$ and $\vec r^\prime$ are along the zig-zag edge. The simulated
results for $\lambda=0.1$ are plotted at different values of $U$ in the bulk
paramagnetic regime. The edge spin correlation exhibits the ferrimagnetic
correlations among $A$ and $B$-sites because the edge breaks the equivalence
between $A$ and $B$-sites. The magnetic correlations are stronger among the
outer $A$-sites, and are weaker among the inner $B$-sites. All of these
correlations obey the power law and their decay exponents ($\alpha$) are
fitted. As further increasing $U$ towards to the bulk antiferromagnetic
regime, the difference between $AA$ and $BB$ correlations become weaker.

Due to the domination of the magnetic correlation at $A$-sites, we use the
decay exponents of $C_{AA}$ to fit the effective Luttinger parameter $K$ for
the helical edge. The three plots in Fig. \ref{fig:edgeaf_power} (a) at $%
U=1,1.5$ and $2$ gives rises to $K=\frac{1}{2}\alpha\approx 0.8, 0.5 $, and $%
0.4$, respectively. The case of $U=1$ belongs to the topological band
insulating phase in which interaction effects are perturbative. For the case
of $U=2$ at which the bulk remains non-magnetic, although the edge remains
gapless, it is only because the conservation of $S_z$ which is \emph{not} an
essential symmetry of topological insulators. As long as the above Umklapp
term Eq. \ref{eq:2pctbg} is introduced, which unfortunately cannot be
simulated by our QMC method, the gapless helical edge states are
destabilized. We argue that the system enters a new phase with paramagnetic
bulk but unstable edges. The transition point between these two paramagnetic
phases at $\lambda=0.1$ lies at $U\approx 1.5$ with $K\approx 0.5$.

We have calculated the edge spin correlations for other values of spin-orbit
coupling and interaction parameters to map the boundary with $K=0.5$ between
two different bulk paramagnetic phases. The boundary is plotted in Fig. \ref%
{fig:phase_diagram}. As $\lambda$ decreases, the dispersion of the edge
spectra becomes more flat, and interaction effects go stronger. As a result,
the boundary shifts to lower values of $U$. In particular at $\lambda=0$,
the edge spectra become exactly flat, we expect edge ferromagnetism at
infinitesimal $U$ due to the density of state divergence. Thus the boundary
should pass the origin. In particular, the edge ferromagnetism of graphene
ribbon has been simulated in Ref. \cite{feldner2011}.


\section{Absence of the spin-orbit order in spin liquid phase at $\protect%
\lambda=0$}

\label{sect:spinliquid}

\begin{figure}[tbp]
\centering
{\epsfig{file=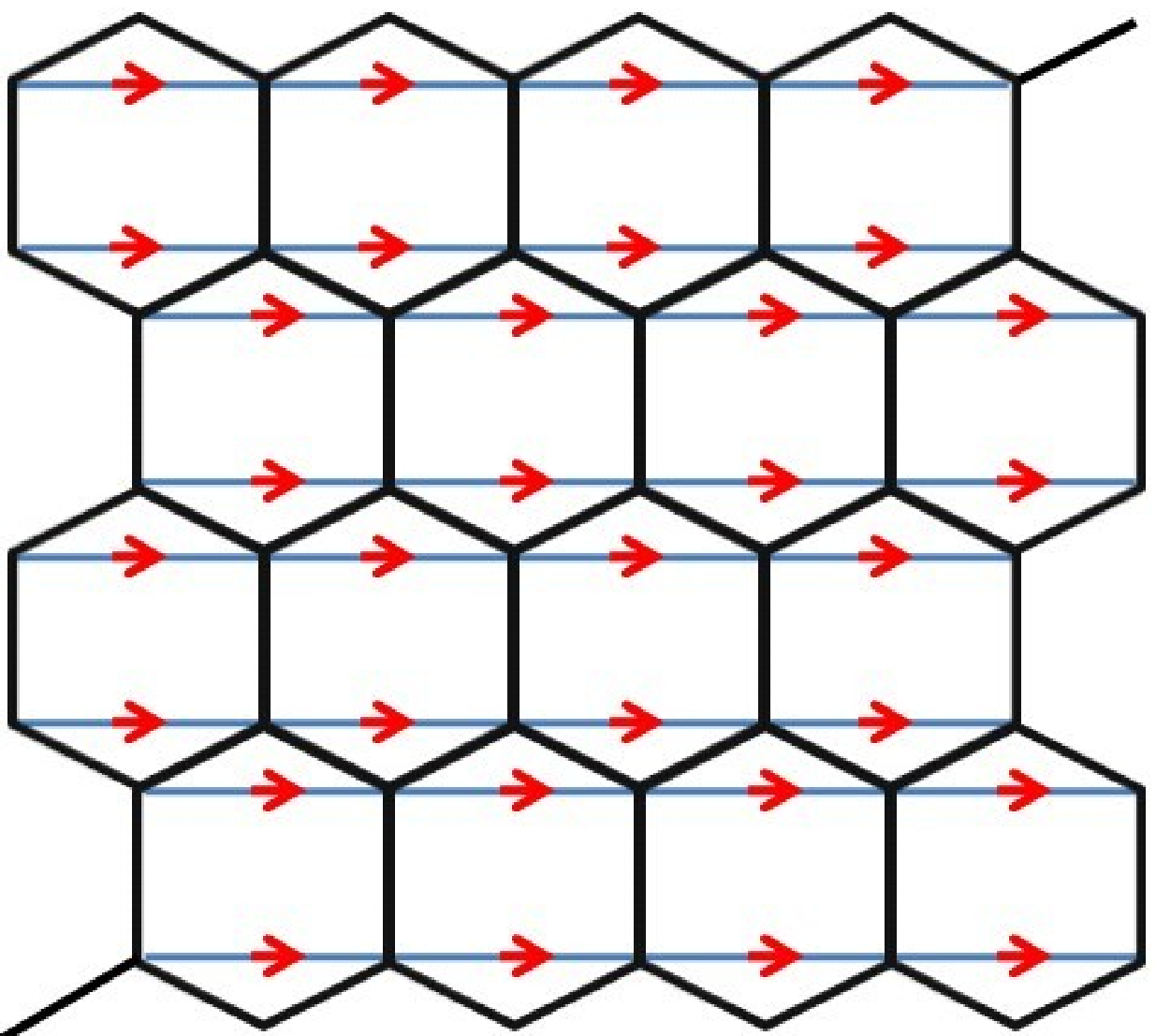,clip=1,width=0.6\linewidth,angle=0}}
\caption{(Color online) The definition of the positive direction for the NNN
bonds in the honeycomb lattice, based on which the NNN current form factors $%
Q_C^{AF}$ and $Q_S^{AF}$ are defined in Eq. \protect\ref{eq:currents}.}
\label{fig:curr_corre}
\end{figure}

Since Meng \textit{et al.} \cite{meng2010} claimed the existence of a
spin-liquid phase for Hubbard model $(\lambda/t=0)$ at $3.7<U/t<4.3$ (see
Fig. \ref{fig:phase_diagram}), it has attracted considerable interests and
debates on the nature of this phase. One possibility of such a phase is that
it could be a relative spin-orbit symmetry breaking phase with a non-trivial
mean-field band structure \cite{kivelson}. If it is the case, a finite $%
\lambda/t$ behaves like an external field to pin down the order parameter
along the external spin-orbit configuration. Then the semi-metal and
spin-liquid phase are indistinguishable at finite $\lambda/t$. In this
section, we will check the form factor of the such a spin-orbit order
parameter between NNN sites at $\lambda=0$, and find negative results.

Without loss of generality, we only consider the horizontal bonds. We define
the positive directions for the NNN horizontal bonds as depicted in Fig. \ref%
{fig:curr_corre}. Two different NNN current orders are designed, including
the charge flux order and the Kane-Mele type spin-orbit order, or,
equivalently, the spin-current flux order. Their form factors are denoted as
$Q_{C}^{AF}$ and $Q_{S}^{AF}$ and are defined as
\begin{eqnarray}
Q_{C}^{AF}&=&\frac{1}{N}\langle G| \big\{ \sum_{i}(-1)^{i}J^{C}_{i,i+\vec{e}%
_x} \big\}^2 |G\rangle,  \notag \\
Q_{S}^{AF}&=&\frac{1}{N}\langle G| \big\{ \sum_{i}(-1)^{i}J^{S}_{i,i+\vec{e}%
_x} \big\}^2 |G\rangle  \label{eq:currents}
\end{eqnarray}
where $(-)^i$ takes the values of $1$ or $-1$ for site $i$ in the $A$ and $B$
sublattices, respectively; the charge current $J^C_{i,i+\vec{e}%
_x}=J^{NNN}_{i,i+\vec{e}_x;\uparrow}+J^{NNN}_{i,i+\vec{e}_x;\downarrow}$,
and spin current $J^S_{i,i+\vec{e}_x}=J^{NNN}_{i,i+\vec{e}_x;\uparrow}
-J^{NNN}_{i,i+\vec{e}_x;\downarrow}$; $\vec{e}_x$ is the NNN vector along
horizontal direction. Please note that the bond current operator here $%
J^{NNN}_{i,i+\vec e_x; \sigma}$ is different from that in Eq. \ref%
{eq:nnn_curr} as
\begin{eqnarray}
J^{NNN}_{i,i+\vec e_x; \sigma}= i \big\{ c^\dagger_{i,\sigma} c_{i+\vec
e_x,\sigma} -h.c.\big\},
\end{eqnarray}
where no summation over $\sigma$ is assumed.

\begin{figure}[tbp]
\centering
{\epsfig{file=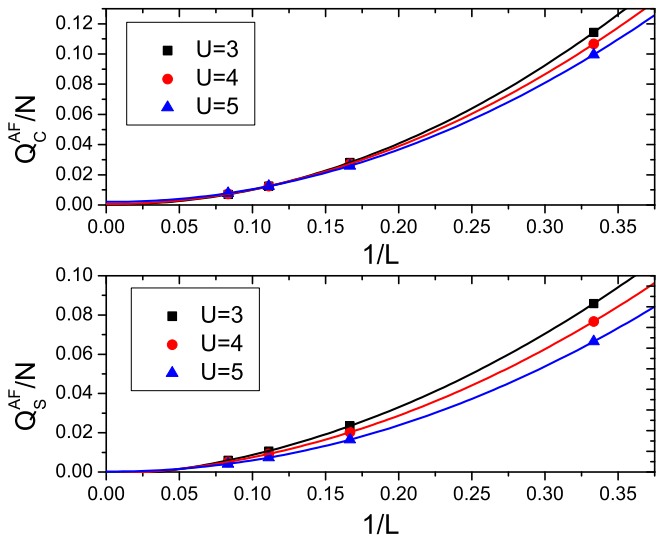,clip=1,width=\linewidth,angle=0}}
\caption{(Color online) The finite-size scaling of the form factors of the
NNN currents in the charge sector $Q_c^{AF}$ and $Q_S^{AF}$ in the spin
sector as defined in Eq. \protect\ref{eq:currents}. The periodical boundary
condition is employed. The sample size is $N=2\times L \times L$ with $%
L=3,6,9$ and $12$. $\protect\lambda$ is set 0 in this calculation, and $U$%
=3,4 and 5.}
\label{fig:nnn_sf}
\end{figure}

We have performed the simulation of the NNN charge and spin-current form
factors defined in Eq. \ref{eq:currents} for the Hubbard model at $\lambda=0$%
. The extrapolations of the form factors to the infinite lattice size are
depicted in Fig. \ref{fig:nnn_sf}. The curves represents three typical
Hubbard $U$ values $U=3,4$ and $5$, which fall in semi-metal phase,
spin-liquid phase and Mott insulating phase, respectively. For all the three
parameters, both the charge and spin NNN current antiferromagnetic form
factors vanish in thermodynamic limit, indicating the absence of the NNN
charge and spin-current orders in all these three phases, especially the
spin liquid phase. The nature of this spin-liquid phase, whether it is
actually a subtly ordered phase or a genuinely exotic phase with non-trivial
topological property, remains an unsolved question.


\section{Conclusions}

\label{sect:conclusion} 
We have studied the particle-hole symmetry in the
KMH model, which results in the absence of the charge and spin currents and
the absence of the quantum Monte-Carlo sign problem. The determinant QMC
simulations have been performed for both the bulk and edge properties. The
bulk antiferromagnetic long range order appears at large values of $U$. With
the open boundary condition, the antiferromagnetic correlation is strongest
along edges. 

We also studied the stability of helical edges in
paramagnetic insulating phase when turn on infinitesimal two-particle
backscattering term, which can be introduced by time-reversal invariant but 
$S_z$ not conserved interaction terms. The paramagnetic
insulating phase in Fig. \ref{fig:phase_diagram} can be classified into two
regimes of weak and intermediate interactions, respectively. In the weak
interaction regime, the helical edge states remain gapless which is robust
against the two-particle back-scattering; in the intermediate interaction
regime, the edge states can spontaneously break time-reversal symmetry by
developing magnetic ordering along the edge by the two-particle
backscattering term. Since this destabilizing helical edges occurs when the
bulk remains time-reversal invariant, it is an interesting and open question
whether the non-trivial bulk $Z_2$-topology is still maintained in this
regime.

We also checked that the spin-liquid phase in the Hubbard model at $%
\lambda=0 $ in the honeycomb lattice is neither a spontaneously developed
Haldane-type quantum anomalous Hall insulator, nor, the Kane-Mele type
quantum spin Hall insulator.


\section*{Acknowledgement}

C. W. and D. Z. thank J. E. Hirsch for his education on the determinate
quantum Monte-Carlo simulations and his encouragement on this work. C. W. is
grateful to S. Kivelson for suggesting the simulation in Section \ref%
{sect:spinliquid}, and thanks L. Balents, L. Fu, T. Xiang for helpful
discussions. C. W. is supported by ARO-W911NF0810291 and Sloan Research
Foundation. D. Z and G. M. Z. are supported by NSF of China and the National
Program for Basic Research of MOST-China.

\textit{Note Added}~ During the preparation of this manuscript, we learned
the work on the QMC simulation on the same KMH model by Hohenadler \textit{%
et al} \cite{hohenadler2010}.

\appendix
\section{Absence of the sign problem for the zero temperature QMC simulations}
\label{sect:appendix}

In this appendix, we prove the absence of the sign problem of the KMH 
model at half-filling for zero temperature determinant QMC, which
is essentially an imaginary time projector algorithm.
The explanation to the algorithm can be find in Ref. \cite{assaad2008}.
For readers' convenience, we also give a brief introduction below.

The Hamiltonian composes of free and interaction parts
\bea
H=H_{t}+H_{I}. 
\eea
The free part reads
\bea
H_{t}=\sum_{i,j}c_{i,\sigma }^{\dag }K_{i,j}^{\sigma }c_{j,\sigma },
\label{eq:HK}
\eea
where the kinetic energy matrix kernels $K^{\uparrow}$ and 
$K^{\downarrow}$  of the Kane-Mele model in Eq. \ref{eq:HK}
are given in Section II.
They satisfy the relation of Eq. \ref{eq:nosign}.
The interaction part is
\bea
H_{I}=\frac{U}{2}\sum_{i}(n_{i\uparrow }+n_{i\downarrow }-1)^{2}.
\label{eq:hhint}
\eea

The expectation value of a physical observable operator $\hat{O}$ at 
zero temperature is defined as
\bea
\langle \hat{O}\rangle =\frac{\langle \psi _{0}|\hat{O}|\psi _{0}\rangle }{%
\langle \psi _{0}|\psi _{0}\rangle }=\frac{\langle \psi _{T}|e^{-\Theta H}%
\hat{O}e^{-\Theta H}|\psi _{T}\rangle }{\langle \psi _{T}|e^{-2\Theta
H}|\psi _{T}\rangle }, 
\label{eq:expec}
\eea
where $|\psi _{0}\rangle $ is the ground state; $\Theta $ is a projection
parameter large enough to ensure the trial wavefunction $|\psi
_{T}\rangle $ is projected to the ground state $|\psi _{0}\rangle $.
The discretized HS transformation of the interaction term 
Eq. \ref{eq:hhint} is performed in the density channel
as the same as that in Eq. \ref{eq:HS}.
The imaginary time propagator, i.e., the projection operator, is represented as
\bea
e^{-\Theta H} &=&\sum_{\{l\}}
\Big\{ U_{\{l\}}(\Theta,0)
\prod_{i,p}\gamma _{i,p}(l)e^{-i\eta _{i,p}(l)\sqrt{\Delta \tau \frac{U}{2}}}
\Big\}, \nn \\
U_{\{l\}}(\Theta,0)&=&
\prod_{p=M}^{1}e^{-\Delta \tau
\sum_{i,j}c_{i\uparrow }^{\dag }K_{ij}^{\uparrow }c_{j\uparrow }}e^{i\sqrt{
\Delta \tau \frac{U}{2}}\sum_{i}c_{i\uparrow }^{\dag }
\eta _{i,p}(l)c_{i\uparrow}}\nn \\
&\times&
\prod_{p=M}^{1}e^{-\Delta \tau \sum_{i,j}c_{i\downarrow
}^{\dag }K_{ij}^{\downarrow }c_{j\downarrow }}e^{i\sqrt{\Delta \tau \frac{U}{2}}
\sum_{i}c_{i\downarrow }^{\dag }\eta _{i,p}(l)c_{i\downarrow }}, \nn \\
\label{eq:imgprop}
\eea
where 
$\gamma_{i,p}(l)$ and $\eta_{i,p}(l)$ are the space-time discretized 
HS fields defined in Eq. \ref{eq:HS} with $l$ taking values of $\pm 1,\pm 2$; 
$\sum_{\{l\}}$ represents the summation over the spatial and temporal 
configurations of the HS field;
$U_{\{l\}}(\Theta,0)$ is the propagation operator for the HS configuration
$\{ l \}$.

The trial wavefunction $|\psi _{T}\rangle $ is required to be a Slater 
determinant, which we will specify later. 
The ground state  $|\psi_0\rangle$ can be obtained from applying the imaginary 
time propagator $e^{-\Theta H}$ of Eq. 
\ref{eq:imgprop} on $|\psi_T\rangle$ as
\bea
|\psi_{0}\rangle 
=
\sum_{\{l\}} \Big\{ U_{\{l\}}(\Theta,0)
\prod_{i,p}\gamma _{i,p}(l)e^{-i\eta _{i,p}(l)\sqrt{\Delta \tau \frac{U}{2}}}
\Big \} |\psi_T\rangle, ~~~
\eea
We further perform the calculation of Eq.~ \ref{eq:expec} as
\begin{widetext}
\bea
\langle \hat{O} \rangle &=& 
\frac{\sum_{\{l\}} \Big\{ \langle \psi _{T} ~|U_{l}
(2\Theta ,\Theta ) ~ \hat{O} ~U_{l}(\Theta, 0)
|\psi _{T}\rangle \prod_{i,p}\gamma
_{i,p}(l)e^{-i\eta _{i,p}(l)\sqrt{\Delta \tau \frac{U}{2}}}\Big\} }{
\sum_{\{l\}} \langle \psi _{T}|U_{l}(2\Theta ,0)|\psi _{T}\rangle 
\prod_{i,p}\gamma _{i,p}(l)e^{-i\eta _{i,p}(l)\sqrt{\Delta \tau \frac{U}{2}}}
}
=\sum_{\{l\}}P_{\{l\}} ~ \langle\hat{O}\rangle _{\{l\}},
\eea
\end{widetext}
where $\avg{O}_{\{l\}}$ is the average value of $\hat O$ for the space-time 
HS configuration $\{ l \}$ defined as
\bea
\avg{O}_{\{l\}}=\frac{\langle \psi
_{T}|U_{\{l\}}(2\Theta ,\Theta ) ~\hat{O}~ U_{\{l\}}(\Theta ,0)|\psi _{T}\rangle }
{\langle \psi _{T}|U_{\{l\}}(2\Theta ,0)|\psi _{T}\rangle },
\eea
and $P_{\{l\} }$ is the corresponding probability of 
the HS field configuration $\{ l \}$ as
\bea
P_{\{l\} }&=&\frac{1}{Z}
 ~\langle \psi _{T}|U_{l}(2\Theta ,0)|\psi _{T}\rangle
\prod_{i,p}\gamma _{i,p}(l)e^{-i\eta _{i,p}(l)\sqrt{\Delta \tau \frac{U}{2}}
}. \nn \\
\eea
$Z$ is defined as $Z=\sum_{\{l\}} P_{\{l\}}$.
The summation over the HS configurations $\{l\}$ can be done by using the Monte
Carlo method.

Next we prove the absence of the sign problem for the KMH model
with purely imaginary NNN hoppings at half-filling in the zero
temperature QMC method, {\it i.e.}, the probability $P_{\{l\}}$ 
is positive-definite.
We factorize the $|\psi_T\rangle=|\psi_T^{N_\uparrow}\rangle
\otimes |\psi_T^{N_\downarrow}\rangle$,
where $|\psi_T^{N^\uparrow}\rangle$ is a Slater-determinant state
for spin-$\uparrow$ electrons with the particle number $N^\uparrow$,
and similar convention applies for $|\psi^{N^\downarrow}_T\rangle$.
Then  $P_{\{l\}}$ reads as
\begin{widetext}
\bea
P_{\{l\}}
&=&\frac{1}{Z} \langle \psi^{N^\uparrow}_T|
\prod_{p=2M}^{1}e^{-\Delta \tau
\sum_{i,j}c_{i\uparrow }^{\dag }K_{ij}^{\uparrow }c_{j\uparrow }}e^{i\sqrt{
\Delta \tau U/2}\sum_{i}\eta _{i,p}(l)( c_{i\uparrow }^{\dag
}c_{i\uparrow }-\frac{1}{2}) } ~|\psi^{N^\uparrow}_T\rangle
\nn \\
&\times& 
\langle \psi^{N^\downarrow}_T|
\prod_{p=2M}^{1}e^{-\Delta \tau
\sum_{i,j}c_{i\downarrow }^{\dag }K_{ij}^{\downarrow }c_{j\downarrow }}e^{i
\sqrt{\Delta \tau U/2}\sum_{i}\eta _{i,p}(l)( c_{i\downarrow }^{\dag
}c_{i\downarrow }-\frac{1}{2})} 
~|\psi^{N^\downarrow}_T\rangle \prod_{i,p}\gamma _{i,p}(l),
\eea
\end{widetext}
where the matrices $K_{ij}^\uparrow$ and $K_{ij}^\downarrow$ satisfy 
the relation of Eq. \ref{eq:nosign};
the HS fields $\gamma _{i,p}(l)$ are positive-definite. 

Let us perform a particle-hole transformation only to 
the spin-$\downarrow$ component
\bea
c_{i\downarrow }^{\dag }\rightarrow d_{i\downarrow }=(-1)^{i}c_{i\downarrow
}^{\dag }, \ \ \, c_{i\downarrow }\rightarrow d_{i\downarrow }^{\dag
}=(-1)^{i}c_{i\downarrow },
\eea
then Slater-determinant state  $|\psi_T^{N^\downarrow}\rangle$
changes to another Slater-determinant state of holes
with the hole number $N-N^\downarrow$
denoted as $|\psi_T^{h,N-N^\downarrow}\rangle$.
We arrive at
\bea
P_{\{l\}}
&=&\frac{1}{Z} \langle \psi_T^{N^\uparrow}|~
\prod_{p=2M}^{1}e^{-\Delta \tau
\sum_{i,j}c_{i\uparrow }^{\dag }K_{ij}^{\uparrow }c_{j\uparrow }}\nn \\
&\times&
e^{i\sqrt{
\Delta \tau U/2}\sum_{i}\eta _{i,p}(l)( c_{i\uparrow }^{\dag
}c_{i\uparrow }-\frac{1}{2}) }~ |\psi_T^{N^\uparrow}\rangle
\nn \\
&\times& 
\langle \psi_T^{h,N-N^\downarrow}|~
\prod_{p=2M}^{1}e^{-\Delta \tau
\sum_{i,j}d_{i\downarrow }^{\dag }K_{ij}^{\downarrow }d_{j\downarrow }}\nn \\
&\times&
e^{-i\sqrt{\Delta \tau U/2}\sum_{i}\eta _{i,p}(l)( d_{i\downarrow }^{\dag
}d_{i\downarrow }-\frac{1}{2})} 
~|\psi_T^{h,N-N^\downarrow}\rangle \nn \\
&\times&
\prod_{i,p}\gamma _{i,p}(l).
\eea

Now we add back the explicit form of the Slater-determinant states
$|\psi_T^{N^\uparrow}\rangle$ and $|\psi_T^{h,N-N^\downarrow}\rangle$ as
\bea
|\psi_T^{N^\uparrow}\rangle&=&\prod_{j=1}^{N^{\uparrow}}
\Big(\sum_{i=1}^{N}c_{i}^{\dag} Q^\uparrow_{i,j} \Big) |0\rangle 
=\prod_{j=1}^{N^\uparrow}\Big(\vec{c}^{\dag }Q^\uparrow \Big)_{j}|0\rangle, \nn\\
|\psi_T^{h,N-N^\downarrow}\rangle&=&
\prod_{j=1}^{N-N^{\downarrow}}\Big(\sum_{i=1}^{N} d_{i}^{\dag} Q_{ij}^\downarrow \Big) 
|0\rangle_h \nn \\
&=&\prod_{j=1}^{N-N^\downarrow}\Big(\vec{d}^{\dag }Q^\downarrow\Big)_{j}|0\rangle_h,
\eea
where $|0\rangle$ and $|0\rangle_h$ are the particle vacuum and hole 
vacuum states, respectively;
$N$ is the number of of lattice sites;
$Q^\uparrow$ is a $N\times N^\uparrow$-dimensional rectangular matrix,
and $Q^\downarrow$ is a $N\times (N-N^\downarrow)$-dimensional matrix;
$\vec c^\dagger$ and $\vec d^\dagger$ are vector notations for $c^\dagger_i$
and $d^\dagger_i$ with $i=1$ to $N$.

The Slater-determinant wavefunction has nice properties as
\bea
e^{\vec c^{\dag} M \vec c} \prod_{j=1}^{N_p}(
\vec c^{\dag }Q)_j|0\rangle
=\prod_{j=1}^{N_{p}} [\vec c^{\dag }e^{M} Q^\uparrow]_j|0 \rangle,
\eea
and
\bea
&&\langle 0|\prod_{j=1}^{N_p}(\vec c Q^\dag)_j
~ e^{\vec c^{\dag} M \vec c} ~
\prod_{j=1}^{N_p}(\vec c^{\dag }Q^\prime)_j|0\rangle\nn \\
&=&\det \left[ Q^\dagger e^{M} Q^\prime \right], 
\eea
where $M$ is an $N\times N$ Hermitian matrix, or anti-Hermitian
matrix.
Based on these properties, we have
\bea
P_{\{l \}}&=&
\det \left [ \left( Q_{\uparrow }\right) ^{\dag }\left(
\prod_{p=2M}^{1}e^{-K^{\uparrow }}e^{iV_{p}(l)}\right) Q_{\uparrow }\right]
\nn \\
&\times& \det\left[\left( Q_{\downarrow }\right) ^{\dag }\left(
\prod_{p=2M}^{1}e^{-K^{\downarrow }}e^{-iV_{p}(l)}\right) Q_{\downarrow
}\right] \nn \\
&\times& \prod_{i,p}\gamma _{i,p}(l),
\eea
where the matrix kernels satisfy $K^\uparrow = (K^\downarrow)^*$
and
$V_{p}(l)$ is a purely real diagonal matrix whose $i$-th 
diagonal element reads 
\bea
[V_{p}(l)]_{ii}=\sqrt{\Delta \tau \frac{U}{2}} ~\eta _{i,p}(l).
\eea

If we set the trial wavefunction to satisfy $N^{\uparrow}=N^{\downarrow}
=N/2$ and $Q^{\downarrow }=(Q^{\uparrow })^{\ast }$, 
then we have
\bea
P_{\{l\}}&=&\frac{1}{Z} \left \vert \det \left[ \left( Q^{\uparrow }\right) 
^{\dag }\left(
\prod_{p=2M}^{1}e^{-K^{\uparrow }}e^{V_{p}(l)}\right) Q^{\uparrow
}\right ]  \right\vert ^{2} \nn \\
&\times&
\prod_{i,p}\gamma _{i,p}(l),
\eea
thus the probability distribution $P_{\{l\} }$ is positive-definite
at half-filling.


\end{document}